\documentclass[11pt]{article}
\usepackage{amsfonts}
\usepackage{amsmath}
\usepackage{amssymb}
\usepackage[a4paper,top=3cm,bottom=3cm,left=2cm,right=2.5cm,bindingoffset=5mm]{geometry}
\usepackage{amscd}

\newtheorem{theorem}{Theorem}

\newtheorem{definition}{Definition}
\newtheorem{example}{Example}

\newtheorem{lemma}{Lemma}

\newtheorem{remark}{Remark}

\numberwithin{equation}{section}
\input{tcilatex}
\begin{document}

\title{Invariance and hierarchy-equivalence}
\author{Nicodemo De Vito\thanks{%
Department of Decision Sciences, Bocconi University. \textit{Email}:
nicodemo.devito@unibocconi.it}}
\date{August 9, 2022}
\maketitle

\begin{abstract}
Two type structures are hierarchy-equivalent if they induce the same set of
hierarchies of beliefs. This note shows that the behavioral implications of
\textquotedblleft cautious rationality and common cautious belief in
cautious rationality\textquotedblright \ (Catonini and De Vito 2021) do not
vary across hierarchy-equivalent type structures.
\end{abstract}

\section{Introduction}

The aim of this note is to state and prove an invariance result for the
behavioral implications of some epistemic assumptions. Precisely, say that
two (epistemic) type structures are \textit{hierarchy-equivalent} if they
represent the same set of hierarchies of beliefs. Consider now the following
epistemic assumptions, studied in Catonini and De Vito (2021):

\begin{enumerate}
\item cautious rationality and common cautious belief in cautious
rationality (R$^{\text{c}}$CB$^{\text{c}}$R$^{\text{c}}$);

\item rationality, transparency of cautiousness, and common cautious belief
in both.
\end{enumerate}

Such epistemic assumptions are represented by events (Borel sets) in a type
structure. The main result of this note (Theorem 1) says that the \textit{%
behavioral implications} of the above epistemic assumptions \textit{do not
vary} across hierarchy-equivalent type structures.

This note is organized as follows. Section 2\ introduces an important,
technical result that will be used in the proofs that follow. Section 3\
briefly reviews the formalism of hierarchies of (lexicographic) beliefs and
type structures, and it introduces the concept of hierarchy-equivalence.
Section 4\ states and proves the main result. Section 5\ concludes.

\section{Preliminaries}

All the spaces considered in this paper are assumed to be topological
spaces. A Souslin space is a topological space that is the image of a
complete, separable metric space under a continuous surjection. In
particular, a Polish space (i.e., a topological space which is homeomorphic
to a complete, separable metric space) is Souslin. For any Borel probability
measure $\mu $ on a topological space $X$, we let $\mu ^{\ast }$\ denote the
outer measure induced by $\mu $. Moreover, if $f:X\rightarrow Y$\ is a Borel
map between topological spaces $X$ and $Y$, we let $\mu \circ f^{-1}$\
denote the image measure of $\mu $\ under $f$. With this, we state a
technical result that will be used in the proof of Lemma 2\ in Section 4.

\begin{lemma}
Fix Souslin spaces $X$ and $Y$,\ and a Borel map $f:X\rightarrow Y$. Then,
for every Borel probability measure $\mu $ on $X$, and for every $E\subseteq
Y$,%
\begin{equation*}
\left( \mu \circ f^{-1}\right) ^{\ast }\left( E\right) =\mu ^{\ast }\left(
f^{-1}\left( E\right) \right) \text{.}
\end{equation*}
\end{lemma}

\noindent \textbf{Proof}. Every Borel probability measure on a Souslin space
is Radon, hence perfect (see Bogachev 2007, Theorem 7.4.3\ and Theorem
7.5.10). It follows from Theorem 3.6 in Peskir (1991) (see also
Hoffmann-Jorgensen 2003, and Dudley 2014, Section 3.4) that the Borel map $%
f:X\rightarrow Y$\ is $\mu $-perfect (that is, it satisfies $\left( \mu
\circ f^{-1}\right) ^{\ast }\left( E\right) =\mu ^{\ast }\left( f^{-1}\left(
E\right) \right) $ for all $E\subseteq Y$)\ for every measure $\mu $ on $X$%
.\hfill $\blacksquare $

\begin{remark}
Lemma 1\ can be equivalently stated in terms of inner measure $\mu _{\ast }$%
; that is, under the stated assumptions, $\left( \mu \circ f^{-1}\right)
_{\ast }\left( E\right) =\mu _{\ast }\left( f^{-1}\left( E\right) \right) $\
for every $E\subseteq Y$.
\end{remark}

Given a sequence $\left( X_{n}\right) _{n\in \mathbb{N}}$ of pairwise
disjoint Souslin spaces, the set $X:=\cup _{n\in \mathbb{N}}X_{n}$\ is
endowed with the \textit{direct sum topology},\footnote{%
In this topology, a set $O\subseteq X$\ is open if and only if $O\cap X_{n}$%
\ is open in $X_{n}$ for all $n\in \mathbb{N}$. The assumption that the
spaces $X_{n}$\ are pairwise disjoint is without any loss of generality,
since they can be replaced by a homeomorphic copy, if needed (see Engelking
1989, p. 75).} so that $X$\ is a Souslin space. Moreover, we endow each
finite or countable product of Souslin spaces with the product topology,
hence the product space is Souslin as well.

We let $\mathcal{M}\left( X\right) $\ denote the set of Borel probability
measures on a topological space $X$. The set $\mathcal{M}\left( X\right) $\
is endowed with the \textit{weak*}-topology. So, if $X$ is Souslin, then $%
\mathcal{M}\left( X\right) $\ is also Souslin. We let $\mathcal{N}\left(
X\right) $\ (resp. $\mathcal{N}_{n}\left( X\right) $) denote the set of all
finite (resp. length-$n$) sequences of Borel probability measures on $X$,
that is,%
\begin{eqnarray*}
\mathcal{N}\left( X\right) &:&=\tbigcup_{n\in \mathbb{N}}\mathcal{N}%
_{n}\left( X\right) \\
&:&=\tbigcup_{n\in \mathbb{N}}\left( \mathcal{M}\left( X\right) \right) ^{n}%
\text{.}
\end{eqnarray*}

Each $\overline{\mu }:=\left( \mu ^{1},...,\mu ^{n}\right) \in \mathcal{N}%
\left( X\right) $\ is called \textbf{lexicographic probability system} (%
\textbf{LPS}). In view of our assumptions, the topological space $\mathcal{N}%
\left( X\right) $ is Souslin.

For every Borel probability measure $\mu $\ on $X$, the support of $\mu $,
denoted by $\mathrm{Supp}\mu $, is the smallest closed subset $C\subseteq X$
such that $\mu \left( C\right) =1$. The support of an LPS $\overline{\mu }%
:=\left( \mu ^{1},...,\mu ^{n}\right) \in \mathcal{N}\left( X\right) $\ is
defined as $\mathrm{Supp}\overline{\mu }:=\cup _{l\leq n}\mathrm{Supp}\mu
^{l}$. So, an LPS $\overline{\mu }:=\left( \mu ^{1},...,\mu ^{n}\right) \in 
\mathcal{N}\left( X\right) $\ is of \textbf{full-support} if $\mathrm{Supp}%
\overline{\mu }=X$. We write $\mathcal{N}^{+}\left( X\right) $ for the set
of full-support LPS's.

Fix Souslin spaces $X$\ and $Y$, and a Borel map $f:X\rightarrow Y$. For
each $n\in \mathbb{N}$, the map $\widehat{f}_{\left( n\right) }:\mathcal{N}%
_{n}\left( X\right) \rightarrow \mathcal{N}_{n}\left( Y\right) $ is defined
by%
\begin{equation*}
\left( \mu ^{1},...,\mu ^{n}\right) \mapsto \widehat{f}_{\left( n\right)
}\left( \left( \mu ^{1},...,\mu ^{n}\right) \right) :=\left( \mu ^{k}\circ
f^{-1}\right) _{k\leq n}\text{.}
\end{equation*}%
With this, the map $\widehat{f}:\mathcal{N}\left( X\right) \rightarrow 
\mathcal{N}\left( Y\right) $ defined by%
\begin{equation*}
\widehat{f}\left( \overline{\mu }\right) :=\widehat{f}_{\left( n\right)
}\left( \overline{\mu }\right) \text{, }\overline{\mu }\in \mathcal{N}%
_{n}\left( X\right) \text{,}
\end{equation*}%
is called the \textbf{image LPS map of} $f$. Alternatively put, the map $%
\widehat{f}$ is the \textit{union} of the maps $\left( \widehat{f}_{\left(
n\right) }\right) _{n\in \mathbb{N}}$, and it is Borel measurable.\footnote{%
For details and proofs related to Borel measurability and continuity of the
involved maps, the reader can consult Catonini and De Vito (2018).}

Given Souslin spaces $X$\ and $Y$, we let $\mathrm{Proj}_{X}$\ denote the
canonical projection from $X\times Y$\ onto $X$; in view of our assumption,
the map $\mathrm{Proj}_{X}$\ is continuous. The marginal measure of $\mu \in 
\mathcal{M}\left( X\times Y\right) $ on $X$ is defined by $\mathrm{marg}%
_{X}\mu :=\mu \circ \mathrm{Proj}_{X}^{-1}$. Consequently, the marginal of $%
\overline{\mu }\in \mathcal{N}\left( X\times Y\right) $\ on $X$ is defined
by $\overline{\mathrm{marg}}_{X}\overline{\mu }:=\widehat{\mathrm{Proj}}%
_{X}\left( \overline{\mu }\right) $, and the function $\widehat{\mathrm{Proj}%
}_{X}:\mathcal{N}\left( X\times Y\right) \rightarrow \mathcal{N}\left(
X\right) $ is continuous and surjective.

Finally, for any set $X$, we let $\mathrm{Id}_{X}$ denote the identity map
on $X$, that is, $\mathrm{Id}_{X}\left( x\right) :=x$\ for all $x\in X$.

\section{Hierarchies of lexicographic beliefs and type structures}

Throughout, we consider finite games. A \textbf{finite game} is a structure $%
G:=\left \langle I,(S_{i},\pi _{i})_{i\in I}\right \rangle $, where (a) $I$
is a finite set of players with cardinality $\left \vert I\right \vert \geq 2
$; (b) for each player $i\in I$, $S_{i}$ is a finite, non-empty set of
strategies; and (c) $\pi _{i}:S\rightarrow \mathbb{R}$ is the payoff
function. Each strategy set $S_{i}$\ is given the obvious topology, i.e.,
the discrete topology.

Fix a finite game $G:=\left \langle I,(S_{i},\pi _{i})_{i\in
I}\right
\rangle $. A \emph{type structure} (associated with $G$)
formalizes Harsanyi's (1967-68) implicit approach to model hierarchies of
beliefs.

\begin{definition}
An $\left( S_{i}\right) _{i\in I}$-\textbf{based lexicographic type structure%
} is a structure $\mathcal{T}:=\langle S_{i},T_{i},\beta _{i}\rangle _{i\in
I}$ where

\begin{enumerate}
\item for each $i\in I$, $T_{i}$\ is a Souslin space;

\item for each $i\in I$, the function $\beta _{i}:T_{i}\rightarrow \mathcal{N%
}\left( S_{-i}\times T_{-i}\right) $ is Borel measurable.
\end{enumerate}

We call each space $T_{i}$\  \textbf{type space} and we call each $\beta _{i}$%
\  \textbf{belief map}. Members of type spaces, viz. $t_{i}\in T_{i}$, are
called \textbf{types}. Each element $\left( s_{i},t_{i}\right) _{i\in I}\in
\tprod_{i\in I}\left( S_{i}\times T_{i}\right) $\ is called \textbf{state} (%
\textbf{of the world}).
\end{definition}

In what follows, we will omit the qualifier \textquotedblleft
lexicographic,\textquotedblright \ and simply speak of \textbf{type
structures} when the underlying strategy sets $\left( S_{i}\right) _{i\in I}$%
\ are clear from the context.

Type structures generate a collection of hierarchies of beliefs for each
player. More precisely, to illustrate how a type induce a hierarchy of
beliefs, we first review how the set of hierarchies is defined. For each $%
i\in I$, set $X_{i}^{1}:=S_{-i}$ and recursively,%
\begin{equation*}
X_{i}^{m+1}:=X_{i}^{m}\times \tprod_{j\neq i}\mathcal{N}\left(
X_{j}^{m}\right) \text{.}
\end{equation*}%
The set of all possible hierarchies of beliefs (LPS's) for player $i$ is%
\begin{equation*}
H_{i}^{0}:=\tprod_{m=1}^{\infty }\mathcal{N}\left( X_{i}^{m}\right) \text{.}
\end{equation*}%
Since each strategy set $S_{i}$\ is finite, it follows form standard
arguments (see Catonini and De Vito 2018) that $X_{i}^{m}$\ ($i\in I$, $%
m\geq 1$) and $H_{i}^{0}$\ ($i\in I$) are Polish (hence Souslin) spaces.

Next, fix a type structure $\mathcal{T}:=\langle S_{i},T_{i},\beta
_{i}\rangle _{i\in I}$ associated with the game $G:=\left \langle
I,(S_{i},\pi _{i})_{i\in I}\right \rangle $. We define, for each player $%
i\in I$, a \textbf{hierarchy map} $d_{i}:T_{i}\rightarrow H_{i}^{0}$ which
associates each $t_{i}\in T_{i}$\ with a hierarchy of beliefs. Such map is
defined recursively.

\begin{itemize}
\item ($m=1$) For each $i\in I$ and each $t_{i}\in T_{i}$, define the
first-order hierarchy map $d_{i}^{1}:T_{i}\rightarrow \mathcal{N}\left(
X_{i}^{1}\right) $ as%
\begin{equation*}
d_{i}^{1}\left( t_{i}\right) :=\overline{\mathrm{marg}}_{S_{-i}}\left( \beta
_{i}\left( t_{i}\right) \right) \text{.}
\end{equation*}%
For each $i\in I$, let $d_{-i}^{1}:=\left( d_{j}^{1}\right) _{j\neq
i}:T_{-i}\rightarrow \tprod_{j\neq i}\mathcal{N}\left( X_{j}^{1}\right) $.

Then, for each $i\in I$, define $\rho _{-i}^{1}:S_{-i}\times
T_{-i}\rightarrow X_{i}^{1}$ as%
\begin{equation*}
\rho _{-i}^{1}:=\left( \mathrm{Id}_{S_{-i}},d_{-i}^{1}\right) \text{.}
\end{equation*}%
Since each belief map $\beta _{i}$\ is Borel measurable, standard arguments
(Catonini and De Vito 2018) show that all the maps defined above are Borel
measurable.

\item ($m+1$, $m\geq 1$) Suppose we have already defined, for each $i\in I$,
Borel measurable maps $d_{i}^{m}:T_{i}\rightarrow \mathcal{N}\left(
X_{i}^{m}\right) $\ and $\rho _{-i}^{m}:S_{-i}\times T_{-i}\rightarrow
X_{i}^{m}$. For each $i\in I$ and each $t_{i}\in T_{i}$, define $%
d_{i}^{m+1}:T_{i}\rightarrow \mathcal{N}\left( X_{i}^{m}\right) $ as%
\begin{equation*}
d_{i}^{m+1}\left( t_{i}\right) :=\widehat{\rho }_{-i}^{m}\left( \beta
_{i}\left( t_{i}\right) \right) \text{.}
\end{equation*}%
For each $i\in I$, let $d_{-i}^{m+1}:=\left( d_{j}^{m+1}\right) _{j\neq
i}:T_{-i}\rightarrow \tprod_{j\neq i}\mathcal{N}\left( X_{j}^{m+1}\right) $.

Then, for each $i\in I$, the map $\rho _{-i}^{m+1}:S_{-i}\times
T_{-i}\rightarrow X_{i}^{m+1}$ is defined as%
\begin{equation*}
\rho _{-i}^{m+1}:=\left( \rho _{-i}^{m},d_{-i}^{m+1}\right) \text{.}
\end{equation*}
All the maps defined above are Borel measurable.
\end{itemize}

With this, for each $i\in I$ the map $d_{i}:T_{i}\rightarrow H_{i}^{0}$\ is
defined by%
\begin{equation*}
d_{i}\left( t_{i}\right) :=\left( d_{i}^{1}\left( t_{i}\right)
,d_{i}^{2}\left( t_{i}\right) ,...\right) \text{.}
\end{equation*}

For future reference, we point out the following fact, whose proof can be
found in Catonini and De Vito (2018).

\begin{remark}
Fix a type $t_{i}\in T_{i}$ in a type structure. Then, for each $m\geq 1$,%
\begin{equation*}
\overline{\mathrm{marg}}_{X_{i}^{m}}\left( d_{i}^{m+1}\left( t_{i}\right)
\right) =d_{i}^{m}\left( t_{i}\right) \text{.}
\end{equation*}%
That is, type $t_{i}$ induces a \textbf{coherent} hierarchy of beliefs.
\end{remark}

Next step is to formally define the notion of hierarchy-equivalence for type
structures. To this end, we first provide the definition of hierarchy
morphism.

\begin{definition}
Fix type structures $\mathcal{T}:=\langle S_{i},T_{i},\beta _{i}\rangle
_{i\in I}$\ and $\mathcal{T}^{\circ }:=\langle S_{i},T_{i}^{\circ },\beta
_{i}^{\circ }\rangle _{i\in I}$ associated with a finite game $%
G:=\left
\langle I,(S_{i},\pi _{i})_{i\in I}\right \rangle $. A map $\left(
\varphi _{i}\right) _{i\in I}:T\rightarrow T^{\circ }$\ is a \textbf{%
hierarchy morphism} (\textbf{from} $\mathcal{T}$\  \textbf{to} $\mathcal{T}%
^{\circ }$) if, for all $i\in I$, $t_{i}\in T_{i}$\ and $n\geq 1$,%
\begin{equation*}
d_{i}^{n}\left( t_{i}\right) =d_{i}^{\circ ,n}\left( \varphi _{i}\left(
t_{i}\right) \right) \text{.}
\end{equation*}
\end{definition}

In words, a hierarchy morphism from $\mathcal{T}$\ to $\mathcal{T}^{\circ }$%
\ is a (not necessarily measurable) map which preserves the hierarchies of
beliefs.\footnote{%
A type morphism is a hierarchy morphism; the reverse implication does not
hold (see Friedenberg and Meier 2011).}

\begin{definition}
Fix type structures $\mathcal{T}:=\langle S_{i},T_{i},\beta _{i}\rangle
_{i\in I}$\ and $\mathcal{T}^{\circ }:=\langle S_{i},T_{i}^{\circ },\beta
_{i}^{\circ }\rangle _{i\in I}$ associated with a finite game $%
G:=\left
\langle I,(S_{i},\pi _{i})_{i\in I}\right \rangle $. We say that $%
\mathcal{T} $\ and $\mathcal{T}^{\circ }$\ are \textbf{hierarchy-equivalent}
if there exist hierarchy morphisms $\left( \varphi _{i}\right) _{i\in
I}:T\rightarrow T^{\circ }$ and $\left( \varphi _{i}^{\circ }\right) _{i\in
I}:T^{\circ }\rightarrow T$.
\end{definition}

The following example illustrates the above concepts.

\begin{example}
Consider a finite game between two players, Ann ($a$)\ and Bob ($b$), such
that%
\begin{eqnarray*}
S_{a} &:&=\left \{ s_{a}\right \} \text{,} \\
S_{b} &:&=\left \{ \bar{s}_{b},\hat{s}_{b}\right \} \text{,}
\end{eqnarray*}%
and $\bar{s}_{b}$\ is \textit{strictly dominant} for Bob. Append to this
game a finite type structure $\mathcal{T}=\langle S_{i},T_{i},\beta
_{i}\rangle _{i\in \left \{ a,b\right \} }$ where each type set is a
singleton, that is, $T_{a}:=\left \{ t_{a}\right \} $\ and $T_{b}:=\left \{ 
\bar{t}_{b}\right \} $. For Bob, $\beta _{b}\left( \bar{t}_{b}\right) $\ is
the trivial probability measure on $\left \{ \left( s_{a},t_{a}\right)
\right \} $. Ann's type is associated with the LPS $\beta
_{a}(t_{a}):=\left( \mu _{a}^{1},\mu _{a}^{2},\mu _{a}^{3}\right) $
described in the following table:%
\begin{equation*}
\begin{tabular}{l|ll}
& $\left( \bar{s}_{b},\bar{t}_{b}\right) $ & $\left( \hat{s}_{b},\bar{t}%
_{b}\right) $ \\ \hline
$\mu _{a}^{1}$ & $\  \  \ 1$ & $\  \  \ 0$ \\ 
$\mu _{a}^{2}$ & $\  \  \ 0$ & $\  \  \ 1$ \\ 
$\mu _{a}^{3}$ & $\ 1/2$ & $\ 1/2$%
\end{tabular}%
\end{equation*}%
Clearly, each type in $\mathcal{T}$\ induces a unique hierarchy of beliefs.

Consider now a different type structure $\mathcal{T}^{\circ }:=\langle
S_{i},T_{i}^{\circ },\beta _{i}^{\circ }\rangle _{i\in \left \{ a,b\right \}
}$ where $T_{a}^{\circ }:=\left \{ t_{a}\right \} $\ and $T_{b}^{\circ
}:=\left \{ \bar{t}_{b},\hat{t}_{b}\right \} $, and the belief maps are as
follows. Bob's belief map satisfies $\beta _{b}^{\circ }\left( \bar{t}%
_{b}\right) =\beta _{b}^{\circ }\left( \hat{t}_{b}\right) =\beta _{b}\left( 
\bar{t}_{b}\right) $. Ann's type is associated with the LPS $\beta
_{a}^{\circ }(t_{a}):=\left( \nu _{a}^{1},\nu _{a}^{2},\nu _{a}^{3}\right) $
described in the following table:%
\begin{equation*}
\begin{tabular}{l|llll}
& $\left( \bar{s}_{b},\bar{t}_{b}\right) $ & $\left( \bar{s}_{b},\hat{t}%
_{b}\right) $ & $\left( \hat{s}_{b},\bar{t}_{b}\right) $ & $\left( \hat{s}%
_{b},\hat{t}_{b}\right) $ \\ \hline
$\nu _{a}^{1}$ & $\ 1/2$ & $\ 1/2$ & $\  \  \ 0$ & $\  \  \ 0$ \\ 
$\nu _{a}^{2}$ & $\  \  \ 0$ & $\  \  \ 0$ & $\ 1/2$ & $\ 1/2$ \\ 
$\nu _{a}^{3}$ & $\  \  \ 0$ & $\ 1/2$ & $\ 1/2$ & $\  \  \ 0$%
\end{tabular}%
\end{equation*}%
Note that $\mathcal{T}$\ and $\mathcal{T}^{\circ }$\ are
hierarchy-equivalent. To see this, define the map $\varphi ^{\circ }:=\left(
\varphi _{a}^{\circ },\varphi _{b}^{\circ }\right) :T_{a}^{\circ }\times
T_{b}^{\circ }\rightarrow T_{a}\times T_{b}$\ as follows:%
\begin{eqnarray*}
\varphi _{a}^{\circ }\left( t_{a}\right) &:&=t_{a}\text{,} \\
\varphi _{b}^{\circ }\left( \bar{t}_{b}\right) &=&\varphi _{b}^{\circ
}\left( \hat{t}_{b}\right) :=\bar{t}_{b}\text{.}
\end{eqnarray*}%
It can be checked that $\varphi ^{\ast }$\ is a hierarchy morphism from $%
\mathcal{T}^{\circ }$\ to $\mathcal{T}$. So, the set of belief hierarchies
induced by types in $\mathcal{T}^{\circ }$\ is included in the set of belief
hierarchies induced by types in $\mathcal{T}$. Since each type in $\mathcal{T%
}$\ induces a unique hierarchy of beliefs, the conclusion follows.\hfill $%
\blacklozenge $
\end{example}

\section{The main result\label{Section: main result}}

In this section we state and prove the result of this note. We first review
the epistemic condition of interest (subsection 4.1), then we show that such
conditions are invariant (in terms of behavioral implications) across
hierarchy-equivalent type structures (subsection 4.2).

\subsection{Epistemic events\label{Subsection: epistemic analysis}}

For this subsection, we fix a finite game $G:=\left \langle I,(S_{i},\pi
_{i})_{i\in I}\right \rangle $, and we append to $G$ a type structure $%
\mathcal{T}:=\langle S_{i},T_{i},\beta _{i}\rangle _{i\in I}$.

For any two vectors $x:=\left( x_{l}\right) _{l=1}^{n},y:=\left(
y_{l}\right) _{l=1}^{n}\in \mathbb{R}^{n}$, we write $x\geq _{L}y$ if either
(a) $x_{l}=y_{l}$ for every $l\leq n$, or (b) there exists $m\leq n$ such
that $x_{m}>y_{m}$ and\ $x_{l}=y_{l}$ for every $l<m$; we write $x>_{L}y$\
if condition (b) holds.

\begin{definition}
A strategy $s_{i}\in S_{i}$\ is \textbf{optimal under} $\beta
_{i}(t_{i}):=\left( \mu _{i}^{1},...,\mu _{i}^{n}\right) \in \mathcal{N}%
(S_{-i}\times T_{-i})$\ if, for every $s_{i}^{\prime }\in S_{i}$,%
\begin{equation*}
\left( \pi _{i}(s_{i},\mathrm{marg}_{S_{-i}}\mu _{i}^{l})\right)
_{l=1}^{n}\geq _{L}\left( \pi _{i}(s_{i}^{\prime },\mathrm{marg}_{S_{-i}}\mu
_{i}^{l})\right) _{l=1}^{n}\text{.}
\end{equation*}%
We say that $s_{i}$\ is a \textbf{lexicographic best reply to} $\overline{%
\mathrm{marg}}_{S_{-i}}\beta _{i}(t_{i})$\ if it\ is optimal under $\beta
_{i}(t_{i})$.
\end{definition}

This is the usual definition of optimality for a strategy, but this time
optimality is taken lexicographically.

\begin{definition}
A type $t_{i}\in T_{i}$ is \textbf{cautious} (in $\mathcal{T}$) if $%
\overline{\mathrm{marg}}_{S_{-i}}\beta _{i}(t_{i})\in \mathcal{N}^{+}\left(
S_{-i}\right) $.
\end{definition}

In words, this notion of cautiousness\ requires that the first-order belief
of a type be a full-support LPS.

For strategy-type pairs we define the following notions.

\begin{definition}
Fix a strategy-type pair $\left( s_{i},t_{i}\right) \in S_{i}\times T_{i}$.

\begin{enumerate}
\item Say $\left( s_{i},t_{i}\right) $\ is \textbf{rational} (in $\mathcal{T}
$) if $s_{i}$\ is optimal under $\beta _{i}\left( t_{i}\right) $.

\item Say $\left( s_{i},t_{i}\right) $\ is \textbf{cautiously rational} (in $%
\mathcal{T}$) if it is rational and $t_{i}$ is cautious.
\end{enumerate}
\end{definition}

The following definition is essential.

\begin{definition}
Fix a non-empty event $E\subseteq S_{-i}\times T_{-i}$ and a type $t_{i}\in
T_{i}$ with $\beta _{i}\left( t_{i}\right) :=(\mu _{i}^{1},...,\mu _{i}^{n})$%
. We say that $E$ is \textbf{cautiously believed under} $\beta _{i}\left(
t_{i}\right) $ \textbf{at level} $m\leq n$\ if the following conditions hold:

\begin{description}
\item[(i)] $\mu _{i}^{l}\left( E\right) =1$\ for all $l\leq m$;

\item[(ii)] for every elementary cylinder $\hat{C}_{s_{-i}}:=\left \{
s_{-i}\right \} \times T_{-i}$, if $E\cap \hat{C}_{s_{-i}}\not=\emptyset $\
then $\mu _{i}^{l}\left( E\cap \hat{C}_{s_{-i}}\right) >0$\ for some $l\leq
m $.
\end{description}

We say that $E$ is \textbf{cautiously believed under} $\beta _{i}\left(
t_{i}\right) $ if it is cautiously believed under $\beta _{i}\left(
t_{i}\right) $ at some level\ $m\leq n$.

We say that $t_{i}\in T_{i}$\  \textbf{cautiously believes} $E$\ if $E$\ is
cautiously believed under $\beta _{i}\left( t_{i}\right) $.
\end{definition}

We are now ready to formally state the epistemic conditions of interest.

For each player $i\in I$, we let $R_{i}^{1}$ denote the set of cautiously
rational strategy-type pairs. Let $\mathbf{B}_{i}^{c}:\Sigma _{S_{-i}\times
T_{-i}}\rightarrow \Sigma _{S_{i}\times T_{i}}$\ be the operator\ defined by%
\begin{equation*}
\mathbf{B}_{i}^{c}\left( E_{-i}\right) :=\left \{ \left( s_{i},t_{i}\right)
\in S_{i}\times T_{i}:t_{i}\text{ cautiously believes }E_{-i}\right \} \text{%
, }E_{-i}\in \Sigma _{S_{-i}\times T_{-i}}\text{.}
\end{equation*}%
As shown in Catonini and De Vito (2021), the set $\mathbf{B}_{i}^{c}\left(
E_{-i}\right) $\ is Borel in $S_{i}\times T_{i}$ if $E_{-i}\subseteq
S_{-i}\times T_{-i}$\ is an event; so the operator $\mathbf{B}_{i}^{c}$ is
well-defined.

For each $m\geq 1$, define $R_{i}^{m+1}$\ recursively by%
\begin{equation*}
R_{i}^{m+1}:=R_{i}^{m}\cap \mathbf{B}_{i}^{c}\left( R_{-i}^{m}\right) \text{,%
}
\end{equation*}%
where $R_{-i}^{m}:=\prod_{j\neq i}R_{j}^{m}$. Note that%
\begin{equation*}
R_{i}^{m+1}=R_{i}^{1}\cap \left( \tbigcap_{l\leq m}\mathbf{B}_{i}^{c}\left(
R_{-i}^{l}\right) \right) \text{,}
\end{equation*}%
and each $R_{i}^{m}$\ is Borel in $S_{i}\times T_{i}$ (see Catonini and De
Vito 2021).

We write $R_{i}^{\infty }:=\cap _{m\in \mathbb{N}}R_{i}^{m}$ for each $i\in
I $. If $\left( s_{i},t_{i}\right) _{i\in I}\in \tprod_{i\in I}R_{i}^{m+1}$,
we say that there is \textbf{cautious rationality and} $m$\textbf{th-order
cautious belief in cautious rationality} (\textbf{R}$^{\text{c}}m$\textbf{B}$%
^{\text{c}}$\textbf{R}$^{\text{c}}$) at this state. If $\left(
s_{i},t_{i}\right) _{i\in I}\in \tprod_{i\in I}R_{i}^{\infty }$, we say that
there is \textbf{cautious rationality and} \textbf{common cautious belief in
cautious rationality} (\textbf{R}$^{\text{c}}$\textbf{CB}$^{\text{c}}$%
\textbf{R}$^{\text{c}}$) at this state.

\subsection{The invariance result}

We consider only the epistemic assumption of R$^{\text{c}}$CB$^{\text{c}}$R$%
^{\text{c}}$, since the proof for the different epistemic assumption
mentioned in the introduction is identical.

\begin{theorem}
Fix type structures $\mathcal{T}=\langle S_{i},T_{i},\beta _{i}\rangle
_{i\in I}$\ and $\mathcal{T}^{\circ }=\langle S_{i},T_{i}^{\circ },\beta
_{i}^{\circ }\rangle _{i\in I}$ associated with a finite game $G:=\left
\langle I,(S_{i},\pi _{i})_{i\in I}\right \rangle $. If $\mathcal{T}$\ and $%
\mathcal{T}^{\circ }$\ are hierarchy-equivalent, then, for each $i\in I$\
and for each $m\geq 0$,%
\begin{equation*}
\mathrm{Proj}_{S_{i}}\left( R_{i}^{m}\right) =\mathrm{Proj}_{S_{i}}\left(
R_{i}^{\circ ,m}\right) \text{.}
\end{equation*}
\end{theorem}

The following lemma plays a crucial role for the proof of Theorem 1.

\begin{lemma}
Fix type structures $\mathcal{T}=\langle S_{i},T_{i},\beta _{i}\rangle
_{i\in I}$\ and $\mathcal{T}^{\circ }=\langle S_{i},T_{i}^{\circ },\beta
_{i}^{\circ }\rangle _{i\in I}$ associated with a finite game $G:=\left
\langle I,(S_{i},\pi _{i})_{i\in I}\right \rangle $. If, for each $i\in I$\
and for each $m\geq 1$,

(a) there are $t_{i}\in T_{i}$\ and $t_{i}^{\circ }\in T_{i}^{\circ }$\ such
that $d_{i}^{m}\left( t_{i}\right) =d_{i}^{\circ ,m}\left( t_{i}^{\circ
}\right) $,

(b) $\mathrm{Proj}_{S_{i}}\left( R_{i}^{m-1}\right) =\mathrm{Proj}%
_{S_{i}}\left( R_{i}^{\circ ,m-1}\right) $,

then, for each $s_{i}\in S_{i}$, $\left( s_{i},t_{i}\right) \in R_{i}^{m}$\
only if $\left( s_{i},t_{i}^{\circ }\right) \in R_{i}^{\circ ,m}$.
\end{lemma}

\noindent \textbf{Proof}. We prove by induction on $m\geq 1$\ that, for each 
$i\in I$, the following statements hold:

(i) for each $t_{i}\in T_{i}$\ and $t_{i}^{\circ }\in T_{i}^{\circ }$, if $%
d_{i}^{m}\left( t_{i}\right) =d_{i}^{\circ ,m}\left( t_{i}^{\circ }\right) $%
\ and $\mathrm{Proj}_{S_{i}}\left( R_{i}^{m-1}\right) =\mathrm{Proj}%
_{S_{i}}\left( R_{i}^{\circ ,m-1}\right) $, then, for each $s_{i}\in S_{i}$, 
$\left( s_{i},t_{i}\right) \in R_{i}^{m}$\ only if $\left(
s_{i},t_{i}^{\circ }\right) \in R_{i}^{\circ ,m}$;

(ii) $\left( \rho _{-i}^{\circ ,m+1}\right) ^{-1}\left( \rho
_{-i}^{m+1}\left( R_{-i}^{m}\right) \right) \subseteq R_{-i}^{\circ ,m}$.

This will yield the result.

\textit{Basis step}. The conclusion of part (i) is immediate: indeed, $%
d_{i}^{1}\left( t_{i}\right) =\overline{\mathrm{marg}}_{S_{-i}}\beta
_{i}\left( t_{i}\right) =\overline{\mathrm{marg}}_{S_{-i}}\beta _{i}^{\circ
}\left( t_{i}\right) =d_{i}^{\circ ,1}\left( t_{i}^{\circ }\right) $, and,
by definition, $R_{i}^{0}:=S_{i}\times T_{i}$ and $R_{i}^{\circ
,0}:=S_{i}\times T_{i}^{\circ }$. Hence, $\left( s_{i},t_{i}\right) \in
R_{i}^{1}$\ implies $\left( s_{i},t_{i}^{\circ }\right) \in R_{i}^{\circ ,1}$%
. To show part (ii), fix any $\left( s_{-i},t_{-i}^{\circ }\right) \in
\left( \rho _{-i}^{\circ ,2}\right) ^{-1}\left( \rho _{-i}^{2}\left(
R_{-i}^{1}\right) \right) $. Then there exists $t_{-i}\in T_{-i}$\ such that 
$\left( s_{-i},t_{-i}\right) \in R_{-i}^{1}$\ and $\rho _{-i}^{\circ
,2}\left( \left( s_{-i},t_{-i}^{\circ }\right) \right) =\rho _{-i}^{2}\left(
\left( s_{-i},t_{-i}\right) \right) $. Hence, $d_{-i}^{1}\left( t_{i}\right)
=d_{-i}^{\circ ,1}\left( t_{-i}^{\circ }\right) $\ and so, by part (i)
(established for $m=1$), we obtain $\left( s_{-i},t_{-i}^{\circ }\right) \in
R_{-i}^{\circ ,1}$.

\textit{Inductive step}. Suppose that the result is true for $m\geq 1$. We
show that it is also true for $m+1$. As for part (i), consider $t_{i}\in
T_{i}$, $t_{i}^{\circ }\in T_{i}^{\circ }$\ and $s_{i}\in S_{i}$\ so that%
\begin{eqnarray*}
d_{i}^{m+1}\left( t_{i}\right) &=&d_{i}^{\circ ,m+1}\left( t_{i}^{\circ
}\right) \text{,} \\
\left( s_{i},t_{i}\right) &\in &R_{i}^{m+1}\text{.}
\end{eqnarray*}%
By coherence of belief hierarchies (Remark 2), we have $d_{i}^{m}\left(
t_{i}\right) =d_{i}^{\circ ,m}\left( t_{i}^{\circ }\right) $. Hence, by part
(i) of the inductive hypothesis, $\left( s_{i},t_{i}^{\circ }\right) \in
R_{i}^{\circ ,m}$. Thus, it suffices to show that $t_{i}^{\circ }$\
cautiously believes $R_{-i}^{\circ ,m}$. To this end, set%
\begin{eqnarray*}
\beta _{i}\left( t_{i}\right) &:&=\overline{\nu }_{i}=\left( \nu
_{i}^{1},...,\nu _{i}^{n}\right) \text{,} \\
\beta _{i}\left( t_{i}^{\circ }\right) &:&=\overline{\mu }_{i}=\left( \mu
_{i}^{1},...,\mu _{i}^{n}\right) \text{.}
\end{eqnarray*}%
(Since $d_{i}^{m+1}\left( t_{i}\right) =d_{i}^{\circ ,m+1}\left(
t_{i}^{\circ }\right) $, LPS's $\overline{\nu }_{i}$\ and $\overline{\mu }%
_{i}$\ have the same length.) For each $p\geq 1$, we let%
\begin{equation*}
\overline{\nu }_{i}^{p}:=\left( \nu _{i}^{p,1},...,\nu _{i}^{p,n}\right) 
\text{ \ and \ }\overline{\mu }_{i}^{p}:=\left( \mu _{i}^{p,1},...,\mu
_{i}^{p,n}\right)
\end{equation*}%
\ denote the $p$-th order beliefs induced by $t_{i}$\ and $t_{i}^{\circ }$,
respectively.

Since $\left( s_{i},t_{i}\right) \in R_{i}^{m+1}:=R_{i}^{m}\cap \mathbf{B}%
_{i}^{c}\left( R_{-i}^{m}\right) $, event $R_{-i}^{m}$\ is cautiously
believed under $\beta _{i}\left( t_{i}\right) :=\left( \nu _{i}^{1},...,\nu
_{i}^{n}\right) $\ at some level $k\leq m$. So, by Proposition 2 in Catonini
and De Vito (2021),%
\begin{equation*}
\mathrm{Proj}_{S_{-i}}\left( R_{-i}^{m}\right) =\tbigcup_{l=1}^{k}\text{%
\textrm{Suppmarg}}_{S_{-i}}\nu _{i}^{l}\text{.}
\end{equation*}%
By the inductive hypothesis, $\mathrm{Proj}_{S_{-i}}\left( R_{-i}^{m}\right)
=\mathrm{Proj}_{S_{-i}}\left( R_{-i}^{\circ ,m}\right) $;\ moreover,%
\begin{equation*}
\mathrm{marg}_{S_{-i}}\nu _{i}^{l}=\mathrm{marg}_{S_{-i}}\mu _{i}^{l}
\end{equation*}%
for all $l=1,...,m$ (i.e., the first-order beliefs induced by $\overline{\nu 
}_{i}$\ and $\overline{\mu }_{i}$\ coincide---see again Remark 2). Hence, it
is enough to show that $\mu _{i}^{l}\left( R_{-i}^{\circ ,m}\right) =1$\ for
all $l=1,...,k$.

Note that $\rho _{-i}^{m+1}\left( R_{-i}^{m}\right) $\ is not necessarily a
Borel set. Yet, for all $l=1,...,k$,%
\begin{eqnarray*}
\left( \nu _{i}^{m+1,l}\right) ^{\ast }\left( \rho _{-i}^{m+1}\left(
R_{-i}^{m}\right) \right) &=&\left( \nu _{i}^{l}\circ \left( \rho
_{-i}^{m+1}\right) ^{-1}\right) ^{\ast }\left( \rho _{-i}^{m+1}\left(
R_{-i}^{m}\right) \right) \\
&=&\left( \nu _{i}^{l}\right) ^{\ast }\left( \left( \rho _{-i}^{m+1}\right)
^{-1}\left( \rho _{-i}^{m+1}\left( R_{-i}^{m}\right) \right) \right) \\
&\geq &\left( \nu _{i}^{l}\right) ^{\ast }\left( R_{-i}^{m}\right) \\
&=&\nu _{i}^{l}\left( R_{-i}^{m}\right) \\
&=&1\text{,}
\end{eqnarray*}%
where the first equality holds by definition, the second equality follows
from Lemma 1, the inequality holds by a trivial fact about inverse images of
functions and by monotonicity of outer measures, while the third equality
holds because $R_{-i}^{m}$\ is Borel.

Recall that $\overline{\nu }_{i}^{m+1}=\overline{\mu }_{i}^{m+1}$, since $%
d_{i}^{m+1}\left( t_{i}\right) =d_{i}^{\circ ,m+1}\left( t_{i}^{\circ
}\right) $. Hence, for all $l=1,...,k$,%
\begin{equation*}
\left( \mu _{i}^{m+1,l}\right) ^{\ast }\left( \rho _{-i}^{m+1}\left(
R_{-i}^{m}\right) \right) =\left( \nu _{i}^{m+1,l}\right) ^{\ast }\left(
\rho _{-i}^{m+1}\left( R_{-i}^{m}\right) \right) =1\text{.}
\end{equation*}%
With this, using the fact that $\mu _{i}^{m+1,l}=\mu _{i}^{l}\circ \left(
\rho _{-i}^{\circ ,m+1}\right) ^{-1}$\ and Lemma 1, we obtain, for all $%
l=1,...,k$,%
\begin{equation*}
1=\left( \mu _{i}^{l}\circ \left( \rho _{-i}^{\circ ,m+1}\right)
^{-1}\right) ^{\ast }\left( \rho _{-i}^{m+1}\left( R_{-i}^{m}\right) \right)
=\left( \mu _{i}^{l}\right) ^{\ast }\left( \left( \rho _{-i}^{\circ
,m+1}\right) ^{-1}\left( \rho _{-i}^{m+1}\left( R_{-i}^{m}\right) \right)
\right) \text{.}
\end{equation*}%
By part (ii) of the inductive hypothesis,%
\begin{equation*}
\left( \rho _{-i}^{\circ ,m+1}\right) ^{-1}\left( \rho _{-i}^{m+1}\left(
R_{-i}^{m}\right) \right) \subseteq R_{-i}^{\circ ,m}\text{.}
\end{equation*}%
Hence, for all $l=1,...,k$,%
\begin{eqnarray*}
\mu _{i}^{l}\left( R_{-i}^{\circ ,m}\right) &=&\left( \mu _{i}^{l}\right)
^{\ast }\left( R_{-i}^{\circ ,m}\right) \\
&\geq &\left( \mu _{i}^{l}\right) ^{\ast }\left( \left( \rho _{-i}^{\circ
,m+1}\right) ^{-1}\left( \rho _{-i}^{m+1}\left( R_{-i}^{m}\right) \right)
\right) \\
&=&1\text{,}
\end{eqnarray*}%
where the first equality holds because $R_{-i}^{\circ ,m}$\ is a Borel set,
and the inequality follows from monotonicity of outer measures. This
concludes the proof of the inductive step of part (i).

To prove part (ii), fix any $\left( s_{-i},t_{-i}^{\circ }\right) \in \left(
\rho _{-i}^{\circ ,m+2}\right) ^{-1}\left( \rho _{-i}^{m+2}\left(
R_{-i}^{m+1}\right) \right) $. Then there exists $t_{-i}\in T_{-i}$\ such
that $\left( s_{-i},t_{-i}\right) \in R_{-i}^{m+1}$\ and $\rho _{-i}^{\circ
,m+2}\left( \left( s_{-i},t_{-i}^{\circ }\right) \right) =\rho
_{-i}^{m+2}\left( \left( s_{-i},t_{-i}\right) \right) $. Hence, $%
d_{-i}^{m+1}\left( t_{-i}\right) =d_{-i}^{\circ ,m+1}\left( t_{-i}^{\circ
}\right) $\ and so, by part (i) (established for $m+1$), we obtain $\left(
s_{-i},t_{-i}^{\circ }\right) \in R_{-i}^{\circ ,m+1}$.\hfill $\blacksquare $

\bigskip

Next, the following observation will be useful.

\begin{remark}
Fix type structures $\mathcal{T}=\langle S_{i},T_{i},\beta _{i}\rangle
_{i\in I}$\ and $\mathcal{T}^{\circ }=\langle S_{i},T_{i}^{\circ },\beta
_{i}^{\circ }\rangle _{i\in I}$.\ Suppose that there exists a hierarchy
morphism $\left( \varphi _{i}\right) _{i\in I}:T\rightarrow T^{\circ }$.
Then, for each $i\in I$\ and for each $E_{i}\subseteq S_{i}\times T_{i}$,%
\begin{equation*}
\mathrm{Proj}_{S_{i}}\left( E_{i}\right) =\mathrm{Proj}_{S_{i}}\left( \left( 
\mathrm{Id}_{S_{i}},\varphi _{i}\right) \left( E_{i}\right) \right) \text{.}
\end{equation*}
\end{remark}

\bigskip

\noindent \textbf{Proof of Theorem 1}. Let $\left( \varphi _{i}\right)
_{i\in I}:T\rightarrow T^{\circ }$ (resp. $\left( \varphi _{i}^{\circ
}\right) _{i\in I}:T^{\circ }\rightarrow T$) be a hierarchy morphism from $%
\mathcal{T}$\ to $\mathcal{T}^{\circ }$\ (resp. from $\mathcal{T}^{\circ }$\
to $\mathcal{T}$). We will prove by induction on $m\geq 0$ that, for each $%
i\in I$,%
\begin{eqnarray*}
\left( \mathrm{Id}_{S_{i}},\varphi _{i}\right) \left( R_{i}^{m}\right)
&\subseteq &R_{i}^{\circ ,m}\text{,} \\
\left( \mathrm{Id}_{S_{i}},\varphi _{i}^{\circ }\right) \left( R_{i}^{\circ
,m}\right) &\subseteq &R_{i}^{m}\text{.}
\end{eqnarray*}%
With this, Remark 3 entails%
\begin{eqnarray*}
\mathrm{Proj}_{S_{i}}\left( R_{i}^{m}\right) &=&\mathrm{Proj}_{S_{i}}\left(
\left( \mathrm{Id}_{S_{i}},\varphi _{i}\right) \left( R_{i}^{m}\right)
\right) \subseteq \mathrm{Proj}_{S_{i}}\left( R_{i}^{\circ ,m}\right) \text{,%
} \\
\mathrm{Proj}_{S_{i}}\left( R_{i}^{\circ ,m}\right) &=&\mathrm{Proj}%
_{S_{i}}\left( \left( \mathrm{Id}_{S_{i}},\varphi _{i}\right) \left(
R_{i}^{\circ ,m}\right) \right) \subseteq \mathrm{Proj}_{S_{i}}\left(
R_{i}^{m}\right) \text{,}
\end{eqnarray*}%
hence $\mathrm{Proj}_{S_{i}}\left( R_{i}^{m}\right) =\mathrm{Proj}%
_{S_{i}}\left( R_{i}^{\circ ,m}\right) $, as desired.

\textit{Basis step}. The result is immediate because $R_{i}^{0}:=S_{i}\times
T_{i}$ and $R_{i}^{\circ ,0}:=S_{i}\times T_{i}^{\circ }$\ for each $i\in I$.

\textit{Inductive step}. Suppose that the result is true for $m\geq 0$. By
Remark 3, it follows that, for each $i\in I$,%
\begin{equation}
\mathrm{Proj}_{S_{i}}\left( R_{i}^{m}\right) =\mathrm{Proj}_{S_{i}}\left(
R_{i}^{\circ ,m}\right) \text{.}
\end{equation}%
We now show that the result is true for $m+1$. We will prove only that $%
\left( \mathrm{Id}_{S_{i}},\varphi _{i}\right) \left( R_{i}^{m+1}\right)
\subseteq R_{i}^{\circ ,m+1}$, since the proof for $\left( \mathrm{Id}%
_{S_{i}},\varphi _{i}^{\circ }\right) \left( R_{i}^{\circ ,m+1}\right)
\subseteq R_{i}^{m+1}$ is similar. Consider any $i\in I$\ and $\left(
s_{i},t_{i}\right) \in R_{i}^{m+1}$. By definition of hierarchy morphism,
types $t_{i}\in T_{i}$\ and $\varphi _{i}\left( t_{i}\right) \in
T_{i}^{\circ }$\ induce the same hierarchy, in particular $d_{i}^{m+1}\left(
t_{i}\right) =d_{i}^{\circ ,m+1}\left( \varphi _{i}\left( t_{i}\right)
\right) $. Using the fact that (4.1) holds\ for each $i\in I$, Lemma 2\
yields $\left( s_{i},\varphi _{i}\left( t_{i}\right) \right) \in
R_{i}^{\circ ,m+1}$, as required.\hfill $\blacksquare $

\section{Final remarks\label{Section: final remarks}}

Some remarks on Lemma 2\ are in order. First, the proof of Lemma 2\ follows
the lines of the proof of Lemma D2 in Friedenberg and Keisler (2021). There
are, however, some differences. Friedenberg and Keisler (2021) consider the
epistemic assumptions of \textquotedblleft rationality and common belief of
rationality\textquotedblright \ within a standard (i.e., length-1 LPS's)
type structure formalism. Since belief is a monotone operator, an assumption
like (b) in Lemma 2\ is not needed for the proof of their result.
Furthermore, Lemma D2 in Friedenberg and Keisler (2021) restricts attention
to \textit{finite} type structures---that is, each type set $T_{i}$ is a
finite, discrete set. Hence, the issue of measurability for sets such as $%
\rho _{-i}^{m+1}\left( R_{-i}^{m}\right) $\ (see the inductive step in the
proof of Lemma 2) does not arise in their proof. However, it can be shown
(using the result in Lemma 1,\ as we did in the proof of Lemma 1) that their
result can be extended to arbitrary, not necessarily finite type structures.

Finally, the proof of Theorem 1 can be easily adapted to show that the
behavioral implications of \textquotedblleft rationality and common belief
in rationality\textquotedblright \ do not very across hierarchy-equivalent
type structures. We conjecture (but we have not proved) that an analogous
result holds for the epistemic assumption of \textquotedblleft rationality
and common strong belief in rationality\textquotedblright \ (Battigalli and
Siniscalchi 2002) within the framework of conditional type
structures---i.e., type structures where types map to conditional
probability systems. Such conjecture stems from the fact that, although
formally distinct, the notions of cautious belief and strong belief have
similar properties: they are both non-monotonic, and---more
importantly---they satisfy a monotonicity property for events with the same
behavioral implications. Specifically, consider events $E$ and $F$\ such
that $E\subseteq F$\ and their projections onto strategy sets are equal. It
is known (see Catonini and De Vito 2021) that, in this case, cautious belief
in $E$\ implies cautious belief in $F$. An analogous conclusion also holds
for strong belief.

\end{document}